\documentclass{IEEEtran}

\usepackage{amsmath, amssymb, comment,cite}
\usepackage{graphicx, color, tabu}
\usepackage[latin1]{inputenc}
\usepackage{ bbm, mathtools}
\usepackage{balance}
\usepackage{epstopdf}
\usepackage[margin=15.5mm,top=18mm,bottom=18mm]{geometry}
\usepackage[norelsize, linesnumbered, ruled, lined, boxed, commentsnumbered]{algorithm2e}
\usepackage{cases}

\usepackage[nodisplayskipstretch]{setspace}
\newcounter{mytempeqcounter}

\newcommand{\bigformulatop}[2]{%
	\begin{figure*}[!t]
		\normalsize
		\setcounter{mytempeqcounter}{\value{equation}}
		\setcounter{equation}{#1}
		#2
		
		\setcounter{equation}{\value{mytempeqcounter}}
		\hrulefill
		\vspace*{4pt}
	\end{figure*}
}

\newcommand{\qh}{{\bf h}}

\newcommand{\qs}{{\bf s}}

\newcommand{\qv}{{\bf v}}
\newcommand{\qw}{{\bf w}}
\newcommand{\qx}{{\bf x}}
\newcommand{\qy}{{\bf y}}

\newcommand{\qA}{{\bf A}}

\newcommand{\qF}{{\bf F}}

\newcommand{\qH}{{\bf H}}

\newcommand{\qS}{{\bf S}}
\newcommand{\qT}{{\bf T}}

\newcommand{\qX}{{\bf X}}
\newcommand{\qY}{{\bf Y}}

\newcommand{\dl}{{\mathtt{dl}}}
\newcommand{\SP}{{\mathtt{sp}}}
\newcommand{\EP}{{\mathtt{ep}}}
\newcommand{\Pil}{{\mathtt{che}}}

\newcommand{\ul}{{\mathtt{ul}}}
\newcommand{\Nguard}{{N_{\mathtt{guard}}}}

\newcommand{\taudl}{\omega_{\dl}}

\newcommand{\Sn}{\sigma_n^2}
\newcommand{\diag}{\mathrm{diag}}

\newcommand{\nm}{MN}

\newcommand{\etpq}{{\eta}_{pq}}
\newcommand{\etpqr}{{\eta}_{pq^\prime}}
\newcommand{\etqr}{\eta_{q^\prime}}
\newcommand{\etq}{\eta_{q}}
\newcommand{\Lpqr}{L_{pq^\prime}}

\newcommand{\betpqi}{\beta_{pq,i}}
\newcommand{\betpqri}{\beta_{pq',i}}
\newcommand{\betpq}{\beta_{pq}}
\newcommand{\betpqr}{\beta_{pq^\prime}}

\newcommand{\gampqi}{\gamma_{pq,i}^\Pil}
\newcommand{\gampqisp}{\gamma_{pq,i}^\SP}
\newcommand{\gampqj}{\gamma_{pq,j}^\Pil}

\newcommand{\gampqrj}{\gamma_{pq^\prime,j}^\Pil}

\newcommand{\Ppq}{P_{q}^{\mathtt{Pil}}}
\newcommand{\Puq}{P_{q}^{\mathtt{dt}}}

\newcommand{\rhopq}{\rho_{q}^{\mathtt{Pil}}}
\newcommand{\rhopqp}{\rho_{q'}^{\mathtt{Pil}}}
\newcommand{\rouq}{\rho_{q}^{\mathtt{dt}}}
\newcommand{\rouqr}{\rho_{q'}^{\mathtt{dt}}}

\newcommand{\hmki}{h_{pq,i}}

\newcommand{\tmki}{\tau_{pq,i}}
\newcommand{\nmki}{\nu_{pq,i}}
\newcommand{\Lmk}{L_{pq}}

\newcommand{\Lmkp}{L_{pq'}}
\newcommand{\Bmki}{\beta_{pq,i}}
\newcommand{\Lmki}{\ell_{pq,i}}

\newcommand{\Kmki}{k_{pq,i}}

\newcommand{\kpmki}{\kappa_{pq,i}}

\newcommand{\Kpq}{k_{pq}}
\newcommand{\Lpq}{\ell_{pq}}

\newcommand{\bvrho}{\overline{\boldsymbol{\varrho}}}
\newcommand{\betta}{\overline{\boldsymbol{\eta}}}
\newcommand{\tmin}{t_{\textrm{min}}}
\newcommand{\tmax}{t_{\textrm{max}}}
\newcommand{\Pmax}{P_{\textrm{max}}}

\IEEEoverridecommandlockouts
\allowdisplaybreaks
\graphicspath{{figures/}}

\begin{document}

\title{\fontsize{0.81cm}{1cm}\selectfont  Cell-Free Massive MIMO with OTFS Modulation: Power Control and Resource Allocation }

\author{Mohammadali Mohammadi$^\dag$, Hien Quoc Ngo$^\dag$, and  Michail Matthaiou$^\dag$\\
\small{
$^\dag$Centre for Wireless Innovation (CWI), Queen's University Belfast, U.K.\\
Email:\{m.mohammadi, hien.ngo, m.matthaiou\}@qub.ac.uk
}}\normalsize

\clearpage\maketitle
\thispagestyle{empty}

\begin{abstract}
We consider the downlink of cell-free massive multiple-input multiple-output (MIMO) systems with orthogonal time frequency space (OTFS) modulation. Two pilot-based channel estimation schemes, namely superimposed pilot-based (SP-CHE) and embedded pilot-based channel estimation (EP-CHE), are applied to estimate the channels at the access points (APs). The SP-CHE scheme superimposes low power pilots onto the data symbols in the delay-Doppler domain to avoid the spectral efficiency (SE) loss due to null guard intervals used in the EP-CHE scheme. In the case of SP-CHE scheme, we consider a max-min fairness optimization problem to jointly optimize the per-user pilot/data power allocation coefficients and per-AP power control coefficients. The complicated non-convex problem is then iteratively solved through two decoupled sub-problems. Moreover, a max-min fairness problem is cast for the EP-CHE scheme, where the optimization variables are the per-AP power control coefficients. Numerical results show that the proposed resource allocation approaches provide at most $42$ and $5$-fold increase in the $95\%$-likely per-user SE for the SP-CHE and EP-CHE scheme, respectively, compared with the uniform power control and in correlated shadowing fading channels.

\let\thefootnote\relax\footnotetext{This work was supported by a research grant from the Department for the Economy Northern Ireland under the US-Ireland R\&D Partnership Programme.}
\end{abstract}
\vspace{-1.5em}
\section{Introduction}~\label{Sec:Intro}
The advent of OTFS modulation has paved the way to support wireless communications in high mobility channel scenarios, such as in unmanned aerial vehicle (UAV) communications, high-speed trains, and vehicle-to-vehicle (V2V) communications~\cite{Wei:WC:2021}. In this regard, the research community has recently manifested a growing interest in making OTFS compatible with the existing transmission technologies, such as millimeter wave communication systems~\cite{Hadani:IMS}, non-orthogonal multiple access~\cite{Ding:TCOM:2019}, and massive MIMO~\cite{Liu:JSAC:2020,Muye:JSAC:2021,Raviteja:TVT:2021}. The consolidation of the OTFS with cell-free massive MIMO systems, as an emerging paradigm for the fifth generation (5G) and  beyond 5G (B5G) wireless networks has been studied in~\cite{Mohammad:OTFS}.

Pursuing the objective of providing reliable wireless communications in high mobility environments, OTFS modulation has been introduced by Hadani \emph{et al}~\cite{Hadani:WCNC:2017} as an alternative to orthogonal frequency-division multiplexing (OFDM) modulation. Although OFDM is a commonly adopted modulation waveform in a variety of 4G and 5G standards, it inherently suffers from inter-carrier interference, caused by the impaired orthogonality between the subcarrier components, when there is high Doppler spread
in time-variant channels. OTFS efficiently addresses this challenge by operating in the delay-Doppler (DD) domain rather than in the time-frequency (TF) domain. Specifically, data symbols are multiplexed in the DD domain and transformed to the TF domain via two-dimensional (2D) orthogonal basis functions, covering the entire TF domain. Therefore, OTFS exploits both time and frequency diversity, providing a significant improvement over OFDM by combating the Doppler effect~\cite{Gaudio:TWC:2021}. Moreover, OTFS transforms a time-variant channel into an effective fairly time-invariant channel in the DD domain, which has sparse and stable characteristics. In this context, exploiting the channel sparsity not only facilitates the channel estimation in high Doppler environments, but also offers good trade-offs between the pilot overhead and channel estimation errors~\cite{Raviteja:TVT:2019}.

In cell-free massive MIMO, with the potential to provide significant SE and energy efficiency, a large number of randomly distributed APs serve a relatively small number of users via time-division duplex (TDD)~\cite{Hien:cellfree}. APs can efficiently use simple linear processing schemes for downlink and uplink transmissions and users can rely on large-scale channel statistics for decoding the signals~\cite{Hien:cellfree}. In our recent work~\cite{Mohammad:OTFS}, we have demonstrated that these outstanding aspects of the cell-free massive MIMO are still accessible in high mobility environments, through the use of OTFS.  These results enable the use of computationally efficient and globally optimal algorithms for power control in the OTFS-based cell-free massive MIMO systems to handle the near-far effects and protect the users from strong interference. The question then arises of whether, and by how much, power control is beneficial over the uniform power control, and this question is precisely what motivates our work.

In this paper, we restrict our discussion to the downlink of OTFS-based cell-free massive MIMO with maximum-ratio precoding at the APs. Two low-complexity channel estimation schemes, termed as SP-CHE and EP-CHE are deployed to estimate the channels at the APs. The SP-CHE scheme superimposes low power pilot symbols onto the data symbols to avoid the SE loss due to the null guard intervals in the EP-CHE scheme. Therefore, SP-CHE scheme can naturally serve more users at the same time as compared with EP-CHE~\cite{Mohammad:OTFS}. The main contributions of our work are as follows:

\begin{itemize}
\item A max-min fairness power control problem is formulated to maximize the smallest SE across all users. In the case of the SP-CHE scheme, a complicated non-convex joint optimization problem of pilot/data power allocation and AP power control coefficient design is formulated subject to per-AP and per-user power constraints. A two-layered iterative approach, which combines successive convex approximation (SCA) and the bisection algorithm is then developed to obtain an improved solution at each step.

\item In the case of the EP-CHE scheme, a max-min fairness power control problem is cast where the optimization variables are the per-AP power control coefficients. The problem is reformulated as a second order cone programming (SOCP) and solved via first order methods.

\item Numerical results show that our proposed resource allocation approaches significantly
improve the SE of the system compared to the uniform power control approach.

\end{itemize}

\textit{Notation:} We use bold upper case letters to denote matrices, and bold lower case letters to denote vectors; the superscripts $(\cdot)^*$ and $(\cdot)^\dag$ stand for the conjugate and conjugate-transpose, respectively; $\mathrm{vec}(\cdot)$ denote the vectorization operation; $\mathrm{circ}\{\qx\}$ represents a circulant matrix whose first column is $\qx$; $\diag\{\cdot\}$ returns the diagonal matrix; the matrix $\qF_N=\big(\frac{1}{\sqrt{N}} e^{-j2\pi\frac{k\ell}{N}}\big)_{k,\ell=0,\cdots,N-1}$
denotes the unitary DFT matrix of dimension $N \times N$; $[\qA]_{(i,:)}$, and $[\qA]_{(:,j)}$ denote the $i$th row, and $j$th column of $\qA$, respectively; the operator $\otimes$ denotes the Kronecker product of two matrices; $\mathbb{N}[a,b]$ represents the set of natural numbers ranging from $a$ to $b$; finally, $\mathbb{E}\{\cdot\}$ denotes the statistical expectation.

\vspace{-0.0em}
\section{System Model}~\label{Sec:SysModel}
We consider a TDD-based cell-free massive MIMO system, where $M_a$ single-antenna APs serve $K_u$ single-antenna users.  An OTFS frame is divided into two phases: uplink payload transmission with channel estimation, and downlink payload transmission. Consider an OTFS frame with $M$ sub-carriers having $\Delta f$ (Hz) bandwidth each, and $N_T=N_{\dl}+N_{\ul}$ symbols having $T$ (seconds) symbol duration, of which $N_{\ul}$ symbols are dedicated for uplink data transmission as well as channel estimation and $N_{\dl}$ symbols are used for downlink data transmission. Without loss of any generality, we set $N_{\dl}=N_{\ul}=N$. Therefore, the total bandwidth of the system is $M\Delta f$, whilst $NT$ is the duration of an OTFS block during downlink or uplink transmission.

\emph{OTFS Modulation and Channel Model: } Modulated data symbols of the $q$th user $\{x_q[k,\ell] , k\in \mathbb{N}[0,N-1], \ell\in \mathbb{N}[0,M-1]\}$ are arranged over the DD grid $\Lambda=\left\{\frac{k}{NT}, \frac{\ell}{M\Delta f}\right\}$, where $k$ and $\ell$ represent the delay and Doppler indices, respectively. An inverse symplectic finite Fourier transform (ISFFT) is applied at the OTFS transmitter to convert the set of $\nm$  data symbols $x_q[k,\ell]$ to the TF domain data symbols $X_q[n,m]$, as
\vspace{-0.0em}
\begin{align}~\label{eq:Xtf}
X_q[n,m] = \frac{1}{\sqrt{MN}}\sum_{k=0}^{N-1}\sum_{\ell=0}^{M-1} x_q[k,\ell] e^{j2\pi(\frac{nk}{N}-\frac{m\ell}{M})},
\end{align}
where $n\in\mathbb{N}[0,N-1]$, and $m\in\mathbb{N}[0,M-1]$ denote the time and frequency indices, respectively. We assume that the $q$th user's transmit power in the uplink is $\mathbb{E}\{|x_q[k,\ell]|^2\}=\Puq$. After applying a Heisenberg transform, $X_q[n,m]$ is converted to a time domain signal as
\vspace{-0.0em}
\begin{align}~\label{eq:st}
s_q(t)=\sqrt{{\eta}_q}\sum_{n=0}^{N-1}\sum_{m=0}^{M-1}
X_q[n,m] g_{tx}(t-nT)
e^{j2\pi m \Delta f(t-nT)},\nonumber
\end{align}
where  $0\leq{\eta}_q\leq1$ is the uplink power control coefficient for the $q$th user and $g_{tx}(t)$ is the transmitter pulse of duration $T$. Then, the OTFS frame $s_q(t)$ passes through the doubly selective channel $h_{pq}(\tau,\nu)$ between the $q$th user and the $p$th AP, whose baseband response in the DD domain is given by
$h_{pq}(\tau,\nu)=\sum_{i=1}^{\Lmk} \hmki \delta(\tau-\tmki)\delta(\nu-\nmki$,
where $\Lmk$ denotes the number of paths from the $q$th user to the $p$th AP, $\tmki$, $\nmki$, and  $\hmki$ denote the delay, Doppler shift, and the channel gain, respectively, of the $i$th path of the $q$th user to the $p$th AP~\cite{Raviteja:TWC:2018}. The complex channel gains $\hmki$  for different $(pq, i)$ are independent random variables with $\hmki\sim\mathcal{CN}(0,\Bmki)$. The delay and Doppler shift for the $i$th path are given by $\tmki = \frac{\Lmki}{M\Delta f}$ and $\nmki = \frac{\Kmki+\kpmki}{N T}$, respectively, where $\Lmki \in \mathbb{N}[0, M-1]$ and $\Kmki\in \mathbb{N}[0, N-1]$ are the delay index and Doppler index of the $i$th path, and $\kpmki\in (-0.5,0.5)$ is a fractional Doppler associated with the $i$th path.  Let $\Kpq$ and $\Lpq$ denote the delay and Doppler taps corresponding to the largest delay and Doppler between the $q$th user and the $p$th AP.

\textbf{Uplink payload data transmission:}
In the uplink, all $K_u$ users simultaneously send their data to the APs.
The received signal at the $p$th AP, $r_p(t)$, is processed via a Wigner transform, which is implemented via a receiver filter with an impulse response $g_{rx}(t)$ followed by a sampler, to obtain the received TF domain samples
\vspace{-0.0em}
\begin{align}
Y_p[n,m]=\int r_p(t) g_{rx}(t-nT)
e^{-j2\pi m \Delta f (t-nT)} dt.
\end{align}

Finally, by applying a symplectic finite Fourier transform (SFFT) on the TF domain received symbols $Y_p[n,m]$, the DD domain signal at the $p$th AP is obtained as
\vspace{-0.0em}
\begin{align}~\label{eq:Ydd}
y_p[k,\ell]\!\! =\! \frac{1}{\sqrt{MN}}\!\!\sum_{k=0}^{N-1}\!\sum_{\ell=0}^{M-1}\!\! Y_p[n,m] e^{-j2\pi(\frac{nk}{N}\!-\!\frac{m\ell}{M})} \!+ \! w_p[k,\ell],
\end{align}
where $w_p[k,\ell]\sim\mathcal{CN}(0,\Sn)$ denotes the received additive noise samples in the DD domain.

For the sake of simplicity of presentation and analysis, we consider the vector form representation of the input-output relationship of the OTFS system in the DD domain. Hence, we arrange the DD domain received (transmit) symbols $y_p[k,\ell]$ ($x_q[k,\ell]$) into the 2D received (transmit) symbol matrix $\qY_p$ ($\qX_q$) according to the DD grid, whose $(k,\ell)$th element is $y_p[k,\ell]$ ($x_q[k,\ell]$). Let $\qx_q=\mathrm{vec}(\qX_q)\in \mathbb{C}^{MN\times 1}$ and $\qy_p=\mathrm{vec}(\qY_p)\in \mathbb{C}^{MN\times 1}$ denote the vector forms of the transmitted symbols $\qX_q$ and the received symbols $\qY_p$ in the DD domain, respectively. Therefore, we have
\vspace{-0.0em}
\begin{align}~\label{eq:yAPm:Vect}
\qy_p = \sum_{q=1}^{K_u} \sqrt{\rouq \eta_q}\qH_{pq} \tilde{\qx}_q + \qw_p,
\end{align}
where $\tilde{\qx}_q=\frac{1}{\sqrt{\Puq}}{\qx}_q \in\mathbb{C}^{MN\times 1}$,  $\rouq=\frac{\Puq}{\Sn}$ is the normalized uplink signal-to-noise ratio (SNR), $\Puq$ is the transmit power of user $q$,  $\qw_p\in\mathbb{C}^{MN\times 1}$ is the corresponding noise vector, and $\qH_{pq}\in \mathbb{C}^{MN\times MN}$ is the effective DD domain channel matrix between the $q$th user and $p$th AP, given by~\cite{KWAN:TWC:2021}
\vspace{-0.0em}
\begin{align}~\label{eq:Hpq}
\qH_{pq}
&=
\sum_{i=1}^{\Lmk}
\hmki \qT_{pq}^{(i)},
\end{align}
where $\qT_{pq}^{(i)}=(\textbf{F}_N \otimes \textbf{I}_M)
\boldsymbol{\Pi}^{\Lmki} \boldsymbol{\Delta }^{\Kmki+\kpmki}
(\textbf{F}_N^\dag \otimes \textbf{I}_M)$,  while $\boldsymbol{\Pi} =\mathrm{circ}\{[0,1,0,\ldots,0]^T_{MN\times 1}\}$ denotes a $\nm\times \nm$ permutation matrix and $\boldsymbol{\Delta} = \diag\{z^0,z^1,\ldots,z^{MN-1}\}$ is a diagonal matrix with $z=e^{\frac{j2\pi}{MN}}$.

\textbf{Channel estimation:} We study the performance of the system under the EP-CHE and SP-CHE schemes. Let $\tau_{max}$ and $\nu_{max}$ be the maximum delay and the maximum Doppler spread among all channel paths. Define $\ell_{max} = \tau_{max}M\Delta f=\max_{p,q}\ell_{pq}$ and $k_{max} = \nu_{max}NT=\max_{p,q}k_{pq}$, which indicates that the DD channel responses of the users have a finite support $[0, \ell_{max}]$ along the delay dimension and $[-k_{max}, k_{max}]$ along the Doppler dimension.

For the EP-CHE scheme, consider $\varphi_q$, with $\mathbb{E}\{|\varphi_q|^2\}=\Ppq$ denoting a known pilot symbol for the $q$th user at a specific DD grid location $[k_q, \ell_q]$. Let  $x_{dq}[k,\ell]$ denote the data symbol at grid point $[k,\ell]$, with $\mathbb{E}\{|x_{dq}[k,\ell]|^2\}=\Puq$, and assume that each pilot is surrounded by a guard region of zero symbols. We note that the guard intervals between two adjacent impulses along the Doppler and delay dimension should not be smaller than $2k_{max}$ and $\ell_{max}$, respectively~\cite{Raviteja:TVT:2019}. To avoid a SE loss,  we assume that users are allowed to use dedicated pilot and guard DD grids of each other for data transmission at the cost of increased channel estimation error. The variance of the minimum mean-square error (MMSE) estimate of the channel vector $\qh_{pq}$ entries via the EP-CHE scheme can be expressed as~\eqref{eq:MMSE1:emb} at the top of the next page~\cite{Mohammad:OTFS},
\bigformulatop{5}{
\begin{align}~\label{eq:MMSE1:emb}
&\gamma_{pq,i}^{\EP}
\triangleq
\mathbb{E}\big\{|\hat{h}_{pq,i}^{\EP}|^2\big\}
=
\frac{\rhopq\etq\betpqi^2}
{\rhopq\etq\betpqi\!
	+ \frac{ \etq}{N}\left(\sum_{q'=1}^{K_u}
\rouqr\frac{\etqr}{\etq}
\sum_{i=1}^{\Lpqr}\betpqri
	-\rouq\frac{(4k_{max}+4\hat{k}+1)}{N}\sum_{i=1}^{\Lmk}\betpqi\right)+1},
\end{align}
}
where $\rhopq=\Ppq/\Sn$ is the normalized SNR of each pilot symbol, $\hat{k}$ denotes the additional guard to mitigate the spread due to fractional Doppler and $\hat{k}\in\big\{0,\ldots,\lfloor \frac{N-4k_{max}-1}{4}\rfloor\big\}$.\footnote{Hereafter, we use the superscripts $\EP$ and $\SP$ to denote the EP-CHE and SP-CHE schemes, respectively.} Therefore, during each frame $K_u\leq \left \lfloor \frac{\nm}{\Nguard}\right \rfloor$ users can be supported via the EP-CHE scheme, where $\Nguard=(2\ell_{max}+1)(4k_{max}+4\hat{k}+1)$ denotes the total overhead per each user.

As an alternative and in order to support more number of users, the SP-CHE scheme can be used, at which the data symbol $x_{dq}[k,\ell]$ is superimposed onto the pilot symbol $\psi_{q}[k,\ell]$ in the DD domain as $x_q[k,\ell] = x_{dq}[k,\ell]+\psi_{q}[k,\ell]$. Accordingly, the variance of the MMSE estimate of the channel vector $\qh_{pq}$ entries via the SP-CHE scheme can be expressed as~\eqref{eq:SP:MMSEche} at the top the next page~\cite{Mohammad:OTFS}.
\bigformulatop{6}{
\begin{align}~\label{eq:SP:MMSEche}
\gamma_{pq,i}^{\SP}
\triangleq
         \mathbb{E}\big\{|\hat{h}_{pq,i}^{\SP}|^2\big\}
=
  \frac{ \rhopq \etq \betpqi^2 }
                    {\rhopq \etq \betpqi
                                    +\sum_{q'\neq q}^{K_u}
                                                    \rhopqp \etqr
	                                                              \sum_{i=1}^{\Lpqr}
	                                                                               \betpqri
	+ \sum_{q'=1}^{K_u}
                        \rouqr\etqr
	\sum_{i=1}^{\Lpqr}
	\betpqri +1}.
\end{align}
}

\textbf{Downlink payload data transmission:}
The APs use conjugate precoding to transmit signals to $K_u$ users. Let $\qs_q=\mathrm{vec}(\qS_q)\in\mathbb{C}^{\nm\times 1}$ be the intended signal vector for the $q$th user, where $\qS_q\in\mathbb{C}^{M\times N}$ representing the symbols in the DD domain, whose $(k,\ell)$th element $s_q[k,\ell]$ is the modulated signal in the $k$th Doppler and $\ell$th delay grid, for $ k\in \mathbb{N}[0,N-1], \ell\in \mathbb{N}[0,M-1]$. Therefore, the signal transmitted from the $p$th AP is
\vspace{-0.0em}
\setcounter{equation}{7}
\vspace{-0.2em}
\begin{align}~\label{eq:xqd}
\qx_{d,p}= \sqrt{\rho_d}
\sum_{q=1}^{K_u}
\eta_{pq}^{1/2}\hat{\qH}_{pq}^\dag \qs_q,
\end{align}
where $\rho_d$ is the normalized SNR of each symbol; $\hat{\qH}_{pq}$ denotes the estimated channel matrix between the $q$th user and $p$th AP; $\qs_q\in\mathbb{C}^{MN\times 1}$ is the intended signal vector for the $q$th user; $\eta_{pq}$, $p=1,\ldots,M_a$, $q=1,\ldots,K_u$ are the power control coefficients chosen to satisfy  the  following power constraint at each AP~\cite{Hien:cellfree}
\vspace{-0.2em}
\begin{align}~\label{eq:AP:powcons}
\mathbb{E}\left\{\|\qx_{d,p}\|^2\right\} \leq \rho_d.
\end{align}
\vspace{-0em}
\section{Downlink SE and Resource Allocation}~\label{Sec:SE}
In this section, we first provide a closed-form expression for the downlink SE. Then, we will present a joint AP power control and power allocation between the pilot and data symbols, for the case of the SP-CHE scheme. Moreover, we provide a per-AP power control for the EP-CHE scheme. 

\vspace{-1.2em}
\subsection{Downlink SE Analysis}
We assume that each user has knowledge of the channel statistics but not of the channel realizations~\cite{Hien:cellfree}. The signal received at the $q$th user can be re-arranged to be suitable for detection of the $r$th entry of the received signal in the DD domain with only statistical channel knowledge at the users as
\vspace{-1em}
\begin{align}~\label{eq:zqr}
z_{d,qr}
&=\mathbb{DS}_{q,\dl} . s_{qr}
   +
   \mathbb{BU}_{q,\dl}. s_{qr} +\mathbb{I}_{q,\dl}. s_{qr'}\nonumber\\
   &\hspace{1em}
   +\sum_{q'\neq q}^{K_u}\mathbb{I}_{qq',\dl} .s_{q'r'}
+w_{\dl,qr},
\end{align}
where $s_{qr}=s_{q}[k,\ell]$, with $r=kM+\ell$, $ k\in \mathbb{N}[0,N-1], \ell\in \mathbb{N}[0,M-1]$ denote the zero-mean independent and identically distributed DD domain information symbols to be transmitted to the $q$th user, which satisfies $s_{qr}\sim \mathcal{N}(0,1)$, while
\vspace{-0.2em}
\begin{align*}
\mathbb{DS}_{q,\dl}
&\triangleq
\sqrt{\rho_d}
\mathbb{E}\Big\{
	\!\sum_{p=1}^{M_a}\!
	\eta_{pq}^{1/2}[{\qH}_{pq}]_{(r,:)}
	[\hat{\qH}_{pq}^\dag]_{(:,r)}\!\Big\},\nonumber\\
\mathbb{BU}_{q,\dl}
& \triangleq
 \sqrt{\rho_d}\sum_{p=1}^{M_a}\!
	\eta_{pq}^{1/2}
	[{\qH}_{pq}]_{(r,:)}
	[\hat{\qH}_{pq}^\dag]_{(:,r)}\!-\mathbb{DS}_{q,\dl},\nonumber\\
\mathbb{I}_{q,\dl}& \triangleq
\sqrt{\rho_d}\sum_{p=1}^{M_a}
	\sum_{r'\neq r}^{MN}\eta_{pq}^{1/2}
	[{\qH}_{pq}]_{(r,:)} [\hat{\qH}_{pq}^\dag]_{(:,r')},\nonumber\\
\mathbb{I}_{qq',\dl}& \triangleq
\sqrt{\rho_d}\sum_{p=1}^{M_a}
	\sum_{\substack{r'=1 }}^{MN}\eta_{pq'}^{1/2}
	[{\qH}_{pq}]_{(r,:)} [\hat{\qH}_{pq'}^\dag]_{(:,r')}.
\end{align*}
Then, by treating the sum of the second, third, fourth, and additive white Gaussian noise terms in~\eqref{eq:zqr} as the effective noise and using the worst-case Gaussian noise argument, the achievable downlink SE of the transmission from the APs to the $q$th user for any finite $M_a$ and $K_u$, is provided at~\eqref{eq:Rdq:final} at the top of the next page~\cite{Mohammad:OTFS}, where $\taudl=\big(1-\frac{N_{\ul}}{N_T}\big)$ reflects the fact that, for an OTFS frame of length $MN_T$ symbols, we spend $MN_{\ul}$ symbols for uplink transmission.
\bigformulatop{10}{
\begin{align}~\label{eq:Rdq:final}
\mathtt{SE}_{q,\dl} =\taudl\log_2\left(1+\frac{\rho_{d}\left(\sum_{p=1}^{M_a}	\sum_{i=1}^{\Lmk} \etpq^{1/2}\gampqi \right)^2}
     {\rho_{d}
	\!\sum_{p=1}^{M_a}
	\etpq
	\!\sum_{i=1}^{\Lmk}
	\betpqi
   \Big(\!
	\sum_{j=1}^{\Lmk}\!
	\gampqj \!+\!
    \sum_{q'\neq q}^{K_u}
	\sum_{j=1}^{\Lmkp}\!
	\frac{\etpqr}{\etpq}\gampqrj
    \!\Big)
	\!+\!
	1}\right),
\end{align}
}
\vspace{-1em}
\subsection{Resource Allocation}~\label{Sec:Perf}
We use the max-min optimization criterion for finding the optimal power coefficients $\eta_{pq}$ as well as per-user power allocation between the pilot and data symbols in the case of SP-CHE scheme. At the optimum point, all users get the same rate. Note that in practical systems, each AP has a transmission power limit. We use per antenna power constraints~\eqref{eq:AP:powcons} in the power allocation problem, which can be expressed as~\cite{Mohammad:OTFS}:
\vspace{-0.2em}
\setcounter{equation}{10}
\begin{align}~\label{eq:dlpowercont}
\sum_{q=1}^{K_u}\sum_{i=1}^{\Lmk}
\eta_{pq}\gamma_{pq,i}^{\Pil}\leq 1,\hspace{3em}\Pil\in\{\EP,\SP\}.
\end{align}

We also compare the result with uniform power allocation, where we assume that each AP transmits with full power whilst the power coefficients are only functions of the AP index, $p$. In this case, the power constraint in~\eqref{eq:dlpowercont} holds with equality and the power control coefficients can be directly computed as
\vspace{-0.2em}
\begin{align}~\label{eq:uni:dl}
\eta_{pq} =\bigg({\sum_{q'=1}^{K_u}\sum_{i=1}^{L_{pq}} \gamma_{pq',i}^{\Pil}}\bigg)^{-1}, \quad\forall q=1,\ldots,K_u.
\end{align}

In what follows, we first investigate the joint optimum per-user pilot/data symbol power allocation and per-AP power control design for the SP-CHE scheme. Then, we determine the optimum AP power control coefficients for the EP-CHE scheme.

\subsubsection{SP-CHE scheme} Since the pilot symbols are superimposed onto the data symbols, for a given total power constraint at each user, i.e., $\Pmax$, we can optimally allocate power between data and pilot symbols along with the AP power control coefficients design to maximize the minimum of the downlink SE of all users.  Let $\boldsymbol{\eta}=[\{\eta_{pq}\}: p=1,\ldots,M_a, q=1,\ldots,K_u]$ be a positive vector containing the APs' power control coefficients and $\boldsymbol{\mu}=[{\mu_1}, \ldots, \mu_{K_u}]$ denotes the users' pilot/data symbol power allocation coefficients.  Thus, the max-min fairness power control can be formulated as
\vspace{-0.1em}
\begin{subequations}~\label{eq:opt:SP}
\begin{align}
\mathcal{P}_1: \max_{\{\boldsymbol{\eta},\boldsymbol{\mu}\}}\hspace{0.5em}&\min_{q=1,\ldots,K_u} \mathtt{SINR}_{q,\dl}(\boldsymbol{\eta},\boldsymbol{\mu})
\\
\textrm{s.t}\hspace{0.7em}&\sum_{q=1}^{K_u}
\eta_{pq}\varrho_{pq}\leq 1,~\forall p,\label{eq:opt:SP:b}
\\
&\hspace{0em}\Ppq\!=\!\mu_q\Pmax,\hspace{1em}\Puq\!=\!(1\!-\!\mu_q)\Pmax,~\forall q,\\
& 0<\mu_q<1 ~\forall q,
\hspace{1em}\eta_{pq}\geq 0, ~\forall p, q,
\end{align}
\end{subequations}
where
\vspace{-0.5em}
\begin{align}
\mathtt{SINR_{q,dl}} (\boldsymbol{\eta},\boldsymbol{\mu})=\frac{\rho_{d}\left(\sum_{p=1}^{M_a} \etpq^{1/2}\varrho_{pq}(\mu_q) \right)^2}
     {\rho_{d}
	\!\sum_{p=1}^{M_a}\!
	\betpq
  	\!\sum_{q'=1}^{K_u}\!
	\etpqr\varrho_{pq'}(\mu_{q'})
    \!+\!
	1},\nonumber
\end{align}
with $\varrho_{pq} (\mu_q)\triangleq\sum_{i=1}^{\Lmk} \gamma_{pq,i}^{\SP}$ and $\betpq=\sum_{i=1}^{\Lmk}\betpqi$. Note that this problem is neither convex nor manageable in terms of finding the global optimum solution due to the highly-coupled variables. However, an alternating optimization approach can be developed, where at each step a solution of~\eqref{eq:opt:SP} is obtained in an efficient manner.
We decouple problem $\mathcal{P}_1$ into two sub-problems: the per-AP power control design and the per-user pilot/data symbol power coefficient design. These two sub-problems are alternatively solved as explained in the following.

\emph{Sub-problem $\mathcal{P}_{1-1}$:} This sub-problem solves the per-AP power control problem for a given set of $\boldsymbol{\varrho}=\bvrho$. It can be readily checked that, this problem entails a maximization of a quasiconcave function with linear constraints. Let us define the slack variables $\varsigma_{pq}=\eta_{pq}^{1/2}$ and  $\vartheta_{p}=\sum_{q=1}^{K_u}
\overline{\varrho}_{pq}\varsigma_{pq}^2$.  The sub-problem $\mathcal{P}_{1-1}$ can be reformulated as a standard convex SOCP problem
\vspace{0em}
\begin{subequations}~\label{eq:opt:p1-2:ephi}
\begin{align}
\mathcal{P}_{1-1}: \max_{\{\varsigma_{pq},\vartheta_{p}\},t} \hspace{1em}&t
\\
\textrm{s.t}\hspace{2.5em}&\|\qv_q\|\leq\frac{1}{\sqrt{t}}\sum_{p=1}^{M_a}
\overline{\varrho}_{pq}\varsigma_{pq},~\forall q,
\\
\qquad&\sum_{q=1}^{K_u}
\overline{\varrho}_{pq}\varsigma_{pq}^2\leq \vartheta_{p}^2,~\forall p,\\
&0\leq\vartheta_{p}\leq1,~\forall p,
\hspace{1em} \eta_{pq}\geq 0,~\forall p, q,
\end{align}
\end{subequations}
where $\qv_q = [\sqrt{\beta_{1q}}\vartheta_1,\ldots, \sqrt{\beta_{M_aq}}\vartheta_{M_a}, \frac{1}{\sqrt{\rho_{d}}}]$.  Therefore, the bisection search method is exploited to find the optimal solution of~\eqref{eq:opt:p1-2:ephi}  as follows. First, the upper and lower bounds of the achievable SINR, i.e., $\mathtt{SINR_{q,dl}} (\boldsymbol{\eta},\bvrho)$, are set to $\tmin$ and $\tmax$, respectively, and the initial SINR is chosen as $t = (\tmin+\tmax)/2$. If Problem $\mathcal{P}_{1-1}$ is feasible for a given SINR $t$, then the lower bound  $\tmin$ will be set to $t$ otherwise the upper bound  $\tmax$ will be set to $t$. Then, a new SINR is chosen as $t = (\tmin+\tmax)/2$ for the next iteration. This procedure is continued until the difference between the upper and the lower bounds becomes smaller than a predefined threshold.

\emph{Sub-problem $\mathcal{P}_{1-2}$:} This problem is obtained from $\mathcal{P}_1$ when $\boldsymbol{\eta}$ is fixed, i.e., $\boldsymbol{\eta}$ is given as the solution of problem $\mathcal{P}_{1-1}$. However, the resulting problem is a non-convex problem with respect to (w.r.t.) $\boldsymbol{\mu}$ due to the complicated form of $\gamma_{pq,i}^{\SP}$ in~\eqref{eq:SP:MMSEche}. Therefore, we recast the problem as a convex optimization problem  as follows. We first rewrite $\gamma_{pq,i}^{\SP}$ as
$\gampqisp(\mu_q) = \frac{ \mu_q a_{pq,i}}{ \mu_q b_{pq,i}	+ c_{p}}$ where $a_{pq,i}=\Pmax\etq\betpqi^2$, $b_{pq,i}=\Pmax \etq \left(\betpqi -\betpq\right)$, and $c_{p}=\Pmax \sum_{q'=1}^{K_u} {\etqr}\betpqr +\Sn$. Note that $\gampqisp(\mu_q)$ is a strictly  increasing function of $\mu_q$. Hence, $\varrho_{pq}$ is a strictly  increasing function of $\mu_q$, whose minimum and maximum ranges correspond to $\mu_q=0$ and $\mu_q=1$, respectively. Denoting $\boldsymbol{\varrho} = [ \{\varrho_{pq}\}, q=1,\ldots,K_u, p=1,\ldots,M_a] \in \mathbb{R}^{M_aK_u\times 1}$, sub-problem $\mathcal{P}_{1-2}$ for given $\boldsymbol{\eta}=\betta$ can be expressed as
\vspace{-0.2em}
\begin{subequations}~\label{eq:opt:p1-1:2}
\begin{align}
\mathcal{P}_{1-2}: \max_{\{\boldsymbol{\varrho}\}, t} \hspace{0.5em}&t \\
\textrm{s.t}\hspace{1.0em} &t\leq\mathtt{SINR_{q,dl}} (\betta,\boldsymbol{\varrho})~\forall q,
~\label{eq:opt:p1-1:2b}\\
&\sum_{q=1}^{K_u}
\eta_{pq} \varrho_{pq}\leq 1,~\forall p,\label{eq:opt:p1-1:2c}
\\
& 0\!<\!\varrho_{pq}\!<\!\varrho_{pq}^{\max},~\forall p, q,\label{eq:opt:p1-1:2d}
\end{align}
\end{subequations}
where $t$ is the SINR that all users achieve.
Problem~\eqref{eq:opt:p1-1:2} is still a non-convex problem due to the non-convex constraint~\eqref{eq:opt:p1-1:2b}. In
the following, we invoke SCA to obtain a suboptimal solution of~\eqref{eq:opt:p1-1:2} iteratively. Before proceeding, we first rewrite~\eqref{eq:opt:p1-1:2b} as
\vspace{-0.5em}
\begin{align}~\label{eq:opt:p1-1:3}
     \frac{\left(\sum_{p=1}^{M_a} \etpq^{1/2}\varrho_{pq} \right)^2}
     {t}\geq
	\!\sum_{p=1}^{M_a}\!
	\betpq
  	\!\sum_{q'=1}^{K_u}\!
	\etpqr\varrho_{pq'}
    \!+\!
	\frac{1}{\rho_{d}},~\forall q.
\end{align}
\vspace{0em}
\begin{algorithm} [t]~\label{Alg:bisec}
\SetAlgoLined
 \LinesNumbered
 \textbf{ Initialization:} Set $n=1$ and select initial points $t^{(0)}$ $\boldsymbol{\varrho}^{(0)}$. Compute $x_q^{(0)}=1/\Big(\sum_{p=1}^{M_a}\etpq\varrho_{pq}^{(0)}\Big)^2$. Define a tolerance $\epsilon>0$ and the maximum number of iterations $N_I$.\\
  \textbf{ Iteration} $n$: Solve~\eqref{eq:opt:p1-1:4} with $x_q^{(0)}$ and $t^{(0)}$. Let $\left(\boldsymbol{\varrho}^{(n),*}, t^{(n),*}, x_{q}^{(n),*}\right)$ be the solution. \\
  With $n\geq2$, stop if $|t^{(n-1),*}-t^{(n),*}|<\epsilon$ or $n=N_I$. Otherwise, go to \textbf{Step 4}.\\
  Update $x_{q}^{(n+1)}=x_{q}^{(n),*}$ and $t^{(n+1)}=t^{(n),*}$. Set $n=n+1$ and go to \textbf{Step 2}.
\caption{Iterative algorithm for solving~\eqref{eq:opt:p1-1:2}}
\end{algorithm}
Let $g(x_q, t)=\frac{1}{x_q t}$  denote the left hand side (LHS) of the inequality in~\eqref{eq:opt:p1-1:3} with $x_q=1/\left(\sum_{p=1}^{M_a} \etpq^{1/2}\varrho_{pq} \right)^{2}$. Note that the numerator of $g(x_q, t)$ is a convex function of $\boldsymbol{\varrho}$ and its denominator is a linear function of $t$. Therefore, $g(x_q, t)$  is convex w.r.t. $\{\boldsymbol{\varrho}, t\}$.  In the $n$th iteration of the SCA, for given points $\{x_q^{(n)},t^{(n)}\}$, a global lower bound for $g(x_q, t)$ by applying the first-order Taylor expansion is given by $g(x_q, t) \geq \bar{g}(x_q, t)=\frac{3}{x_{q}^{(n)} t^{(n)}}-\frac{x_q}{\left(x_{q}^{(n)}\right)^2t^{(n)}}-\frac{t}{x_{q}^{(n)} \left(t^{(n)}\right)^2},~\forall q$.
By replacing the  LHS of the~\eqref{eq:opt:p1-1:3} with its lower bound, for any given points $\{x_q^{(n)},t^{(n)}\}$, problem~\eqref{eq:opt:p1-1:2} is approximated as
\vspace{-0.2em}
\begin{subequations}~\label{eq:opt:p1-1:4}
\begin{align}
\mathcal{P}_{1-2}: \max_{\{\boldsymbol{\varrho}, x_q\}, t} \hspace{0.5em}&t \\
\hspace{-2em}\textrm{s.t}\hspace{2.0em} &
\bar{g}(x_q, t)\!\geq\!
	\!\sum_{p=1}^{M_a}\!
	\betpq
  	\!\sum_{q'=1}^{K_u}\!
	\etpqr\varrho_{pq'}
    \!+\!
	\frac{1}{\rho_{d}},~\forall q,\\
&\hspace{-2em}\sum_{q=1}^{K_u}
\eta_{pq} \varrho_{pq}\leq 1,~\forall p,
~0\!<\!\varrho_{pq}\!<\!\varrho_{pq}^{\max},~\forall p, q.\label{eq:opt:p1-1:4d}
\end{align}
\end{subequations}

Problem~\eqref{eq:opt:p1-1:4} is a convex optimization problem, the optimal solution of which can be obtained
using the standard convex program solvers such as CVX~\cite{cvx}. The proposed iterative
algorithm for solving problem~\eqref{eq:opt:p1-1:2} is given in \textbf{Algorithm 1}.

Combing sub-problems $\mathcal{P}_{1-1}$ and $\mathcal{P}_{1-2}$, we can obtain the optimal
solution for $\mathcal{P}_{1}$ as summarized in \textbf{Algorithm 2}.

\subsubsection{EP-CHE scheme}
 We formulate the following max-min fairness power control problem
\vspace{-0.2em}
\begin{subequations}~\label{eq:pot:P2}
\begin{align}
\mathcal{P}_2:  \max_{\boldsymbol{\eta}} \hspace{1em}&\min_{q=1,\ldots,K_u} \mathtt{SINR}_{q,\dl}(\boldsymbol{\eta})\\
\textrm{s.t}\hspace{1em}&\sum_{q=1}^{K_u}\sum_{i=1}^{\Lmk}
	\eta_{pq}\gamma_{pq,i}^{\EP}\leq 1,\quad p=1,\ldots,M_a,\\
\hspace{3em}&\eta_{pq}\geq 0, \quad q=1,\ldots,K_u,~p=1,\ldots,M_a.
\end{align}
\end{subequations}

Problem $\mathcal{P}_{2}$ is a quasiconcave problem. By using similar steps as in~\eqref{eq:opt:p1-2:ephi}, problem $\mathcal{P}_{2}$ can be formulated as a SOCP.  Therefore,  problem $\mathcal{P}_{2}$ can be solved using the bisection method and solving a sequence of convex feasibility problems. 

\vspace{0em}
\begin{algorithm} [t]~\label{Alg:final}
\SetAlgoLined
 \LinesNumbered
  \textbf{Initialization}: set $i=1$. Choose the initial value of $\boldsymbol{\varrho}^{(0)}$, Define a tolerance $\epsilon$ and the maximum number of iterations $N_I$.\\
  \textbf{Iteration} $i:$
    Solve~\eqref{eq:opt:p1-2:ephi} via the \emph{bisection algorithm}~\cite{Boyd} for given $\boldsymbol{\varrho}^{(0)}$. Let $\boldsymbol{\eta}^{(i),*}$ be the solution.\\
    Solve~\eqref{eq:opt:p1-1:4} using \textbf{Algorithm 1}. Let $\boldsymbol{\varrho}^{(i),*}$ be the solution.\\
  Let  $\theta^{(i),*}$ be the optimal value of $\mathcal{P}_1$.\\
  With $i\geq2$, stop if $|\theta^{(i-1),*}-\theta^{(i),*}|<\epsilon$ or $i=N_I$. Otherwise, go to \textbf{Step 6}.\\
  Update $\boldsymbol{\eta}^{(i+1),*}=\boldsymbol{\eta}^{(i),*}$ and $\boldsymbol{\varrho}^{(0)}=\boldsymbol{\varrho}^{(n),*}$. Set $n=n+1$ and go to \textbf{Step 2}.
\caption{Iterative algorithm to solve~\eqref{eq:opt:SP}}
\end{algorithm}

\vspace{-0.2cm}
\subsection{Complexity  and Convergence Analysis}
Here, we provide the computational complexity of Algorithm 2, which involves a SOCP problem in~\eqref{eq:opt:p1-2:ephi} and a feasibility linear problem~\eqref{eq:opt:p1-1:4} at each iteration. For the bisection search method to solve~\eqref{eq:opt:p1-2:ephi} at each iteration, the number of required arithmetic operations is $\mathcal{O}\left( \sqrt{n_{l1} + n_{q1}}(n_{v1} + n_{l1} + n_{q1})n_{v1}^2\right)$, where $n_{v1}=M_aK_u$ is the number of real-valued scalar decision variables, $n_{l1}=M_a$ denotes the number of linear constraints, and $n_{q1}=(M_a+K_u)$ is the number of quadratic constraints~\cite{Vincent:TWC:2017}. Moreover, the total number of the iterations is given by $\log_2\left(\frac{\tmax-\tmin}{\epsilon}\right)$, where $\epsilon$ refers to a predetermined threshold~\cite{Boyd}. Therefore, the per-iteration computational complexity for solving~\eqref{eq:opt:p1-2:ephi}  is $\log_2\left(\frac{\tmax-\tmin}{\epsilon}\right)\mathcal{O}(\sqrt{n_{l1} + n_{q1}}(n_{v1} + n_{l1} + n_{q1})n_{v1}^2)$. Problem~\eqref{eq:opt:p1-1:4} involves $n_{v2}=(M_aK_u+2)$ real-valued scalar decision variables and $n_{l2}=(K_u+M_a+K_uM_a)$ linear constraints. According to~\cite{Vincent:TWC:2017}, the per-iteration cost to solve~\eqref{eq:opt:p1-1:4} is $\mathcal{O}\left((n_{l2}+n_{v2})n_{v2}^{2}n_{l2}^{0.5}\right)$.

The convergence of the objective function in Algorithm 2 is established as follows. We note that an optimal solution of~\eqref{eq:opt:p1-1:4} is also a feasible solution for~\eqref{eq:opt:p1-1:3}, due to the convexity of the bounds used in SCA. Therefore, Algorithm 2 results in a non-increasing sequence of the objectives. On the other hand, since the objective function of~\eqref{eq:opt:p1-1:4} is bounded from below due to the power constraints, Algorithm 2 converges to the global solution.

\vspace{-0.5em}
\section{Numerical Results and Discussions}~\label{Sec:Numer}
We consider an OTFS system with operating carrier frequency $f_c=4$ GHz and the sub-carrier spacing is $\Delta f=15$ kHz. The maximum moving speed in the scenario is $300$ kmph, yielding a maximum Doppler index $k_{max}=9$. We consider the 3GPP vehicular model, extended vehicular A, with $L_{pq}=9$ and  $\tau_{max}=2.5~\mu$sec. We assume that $M_a$ APs and $K_u$ users are uniformly distributed at random within a square of size $1 \times 1~\text{km}^2$ whose edges are wrapped around to avoid the boundary effects. The large-scale fading coefficient models the path loss
and shadow fading, according to $\beta_{pq,i} = \mathrm{PL}_{pq,i} 10 ^{\frac{\sigma_{sh} z_{pq,i}}{10}}$, where $\mathrm{PL}_{pq,i}$ represents the path loss, and $10 ^{\frac{\sigma_{sh} z_{pq,i}}{10}}$ represents the shadow fading with the standard deviation $\sigma_{sh}$, and $z_{pq,i} \sim \mathcal{CN}(0,1)$. We set $\sigma_{sh}=8$ dB and use the three-slop model for the path-loss given in~\cite{Hien:cellfree}. We further use the correlated shadowing model for $d_{pq}>d_1$ as described in~\cite{Hien:cellfree}. We set the noise figure $F = 9$ dB, and thus the noise power $\Sn=-108$ dBm ($\Sn=k_B T_0 (M\Delta f) F$ W, where $k_B$ is the Boltzmann constant, while $T_0=290^o$K is the noise temperature). Let $\tilde{\rho}_d =1$ W, $\Pmax =1$ W be the maximum transmit power of the APs and users, respectively.

\begin{figure}[t]
	\centering
	\vspace{0em}
	\includegraphics[width=97mm, height=63mm]{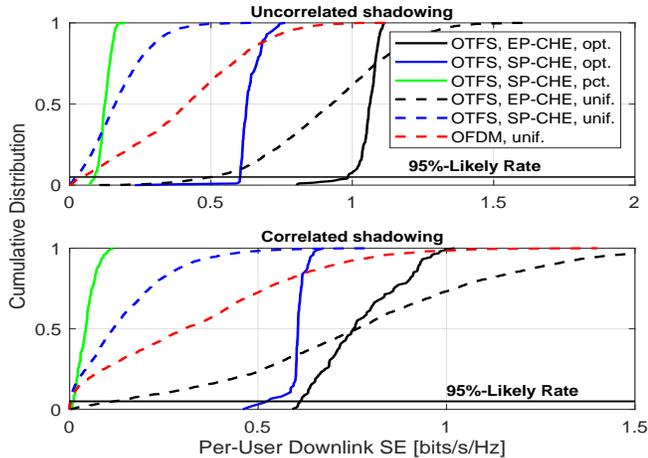}
	\vspace{-1.9em}
	\caption{The cumulative distribution of the per-user downlink throughput for optimum and uniform resource allocation approaches ($M_a=100$, $K_u=20$).}
	\vspace{-1.0em}
	\label{fig:Fig1}
\end{figure}
Fig.~\ref{fig:Fig1}, shows the empirical cumulative distribution functions (CDFs) of the per-user downlink SE for the proposed resource allocation (opt.) and uniform power allocation coefficient (unif.) designs. To quantify the impact of the per-user power allocation between the data and pilot symbols, in the baseline OTFS system with the SP-CHE scheme, the available power budget is equally divided between the pilot and data symbols. Moreover, for the SP-CHE scheme we include the curves for pure AP power control (pct.). It can be observed that, for the EP-CHE and SP-CHE scheme, the optimal resource allocation provides up to $2$ and $20$-fold improvement over the uniform power allocation, respectively. Moreover, while pct. design in the SP-CHE scheme improves the $95\%$ likely rate of the OTFS system in uncorrelated shadow fading channels, it fails to provide satisfactory performance gain in the correlated scenarios, which highlights the importance of joint resource allocation. Finally, the minimum SE of the network improves substantially.

\begin{figure}[t]
	\centering
	\vspace{0em}
	\includegraphics[width=97mm, height=63mm]{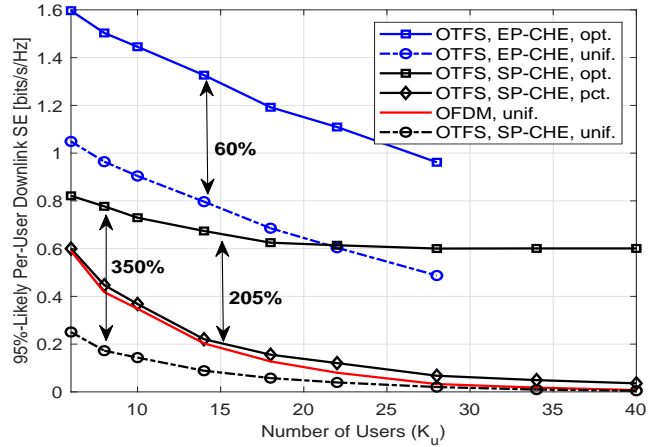}
	\vspace{-1.9em}
	\caption{$95\%$-likely per-user SE for the proposed and uniform resource allocation approaches for an uncorrelated shadowing fading scenario ($M_a=128$).}
	\vspace{-1.0em}
	\label{fig:Fig2}
\end{figure}

Fig.~\ref{fig:Fig2} compare the $95\%$-likely per-user SE of our proposed resource allocation algorithms against $K_u$. The OFDM SE is also illustrated as reference. We observe that SP-CHE with joint resource allocation at the AP and user side approaches the SE of the EP-CHE scheme as soon as $K_u$ is sufficiently large. On the other hand, when $K_u\geq \left \lfloor \frac{\nm}{\Nguard}\right \rfloor = 28$, EP-CHE scheme  can no longer be applied due to the system design constraints, whilst the SP-CHE scheme provides fairly good SE performance for such scenarios.

\vspace{-1.5em}
\section{Conclusion}~\label{Sec:conclusion}
We proposed per-user pilot and data symbols power allocation together with per-AP power control design  in downlink cell-free massive MIMO systems with OTFS modulation to maximize the minimum SE across all users. An alternating optimization algorithm was proposed for solving the non-convex problem. Our results confirmed the significant fairness improvement of the proposed algorithms compared to uniform power control scheme for both SP-CHE and EP-CHE schemes. Moreover, SP-CHE scheme with optimum resource allocation is more robust to correlated shadowing fading as compared to the EP-CHE scheme.

\vspace{-1.5em}

\begin{thebibliography}{1}



\bibitem{Wei:WC:2021}
Z. Wei \emph{et al.,} ``Orthogonal time-frequency space modulation: A promising
next-generation waveform," \emph{IEEE Wireless Commun.,} vol. 28, no. 4, pp.
136-144, Aug. 2021.	

\bibitem{Hadani:IMS}
R. Hadani et al., ``Orthogonal time frequency space (OTFS) modulation
for millimeter-wave communications systems," in \emph{Proc. IEEE MTT-S Int.
Microw. Symp.,} June 2017, pp. 681-683.

\bibitem{Ding:TCOM:2019}
Z. Ding, R. Schober, P. Fan, and H. V. Poor, ``OTFS-NOMA: An efficient
approach for exploiting heterogenous user mobility profiles," \emph{IEEE Trans.
Commun.,} vol. 67, no. 11, pp. 7950-7965, Nov. 2019.


\bibitem{Liu:JSAC:2020}
Y. Liu, S. Zhang, F. Gao, J. Ma, and X. Wang, ``Uplink-aided high mobility downlink channel estimation over massive MIMO-OTFS system,"
\emph{IEEE J. Sel. Areas Commun.,} vol. 38, no. 9, pp. 1994-2009, Sep. 2020.


\bibitem{Muye:JSAC:2021}
M. Li, S. Zhang, F. Gao, P. Fan, and O. A. Dobre, ``A new path division multiple access for the massive MIMO-OTFS networks," \emph{IEEE J. Sel. Areas Commun.,} vol. 39, no. 4, pp. 903-918, Apr. 2021.

\bibitem{Raviteja:TVT:2021}
B. C. Pandey, S. K. Mohammed, P. Raviteja, Y. Hong, and E. Viterbo,
``Low complexity precoding and detection in multi-user massive MIMO
OTFS downlink," \emph{IEEE Trans. Veh. Technol.,} vol. 70, no. 5, pp. 4389-4405, May 2021.


\bibitem{Mohammad:OTFS}
M. Mohammadi, H. Q. Ngo, and M. Matthaiou, ``Cell-free massive MIMO
meets OTFS modulation," \emph{submitted to IEEE Trans. Commun.,} [Online].
Available: https://arxiv.org/abs/2112.10869., 2021.


\bibitem{Hadani:WCNC:2017}
R. Hadani \emph{et al.},  ``Orthogonal time frequency space modulation," in \emph{Proc. IEEE WCNC}, Mar. 2017.

\bibitem{Gaudio:TWC:2021}
L. Gaudio and G. Colavolpe, ``OTFS vs. OFDM in the presence of
sparsity: A fair comparison," \emph{to appear in IEEE Trans. Wireless Commun.,}
2021.

\bibitem{Raviteja:TVT:2019}
P. Raviteja, K. T. Phan, and Y. Hong, ``Embedded pilot-aided channel estimation for OTFS in delay-Doppler channels," \emph{IEEE Trans. Veh. Technol.,} vol. 68, no. 5, pp. 4906-4917, May 2019.


\bibitem{Hien:cellfree}
H. Q. Ngo, A. Ashikhmin, H. Yang, E. G. Larsson, and T. L. Marzetta, ``Cell-free massive MIMO versus small cells," \emph{IEEE Trans. Wireless Commun.,} vol. 16, no. 3, pp. 1834-1850, Mar. 2017.


\bibitem{Raviteja:TWC:2018}
P. Raviteja, K. T. Phan, Y. Hong, and E. Viterbo, ``Interference cancellation and iterative detection for orthogonal time frequency space modulation," \emph{IEEE Trans. Wireless Commun.,} vol. 17, no. 10, pp. 6501-6515, Oct. 2018.


\bibitem{KWAN:TWC:2021}
S. Li, J. Yuan, W. Yuan, Z. Wei, B. Bai, and D. W. K. Ng, ``Performance
analysis of coded OTFS systems over high-mobility channels," \emph{IEEE
Trans. Wireless Commun.,} vol. 20, no. 9, pp. 6033-6048, Sept. 2021.

\bibitem{cvx}
M. Grant and S. Boyd, ``CVX: Matlab software for disciplined convex programming,
version 2.1, [Online]. available: http: //cvxr.com/cvx, 2014." 2014.

\bibitem{Boyd}
S. Boyd, S. P. Boyd, and L. Vandenberghe, \emph{Convex optimization}. Cambridge,
U.K.: Cambridge Univ. Press., 2004.

\bibitem{Vincent:TWC:2017}
H. H. M. Tam, H. D. Tuan, D. T. Ngo, T. Q. Duong, and H. V. Poor, ``Joint
load balancing and interference management for small-cell heterogeneous
networks with limited backhaul capacity," \emph{IEEE Trans. Wireless Commun.,}
vol. 16, no. 2, pp. 872-884, Feb. 2017.

\end{thebibliography}

\bibliographystyle{IEEEtran}

\end{document}